\documentclass[a4paper, draftcls, onecolumn]{IEEEtran}
\usepackage{cite, hyperref}
\usepackage{amsmath,amssymb,amsfonts, multirow}
\usepackage{algorithmic, booktabs}
\usepackage{graphicx}
\usepackage{textcomp}
\usepackage{xcolor}
\def\BibTeX{{\rm B\kern-.05em{\sc i\kern-.025em b}\kern-.08em
    T\kern-.1667em\lower.7ex\hbox{E}\kern-.125emX}}

\begin{document}

\title{Uplink Grant-Free Random Access Solutions for URLLC services in 5G New Radio
}

\author{
Nurul Huda Mahmood{\small $^{1}$}\thanks{This work has been submitted to the IEEE for possible publication. Copyright may be transferred without notice, 
after which this version may no longer be accessible.},
Renato Abreu{\small $^{2}$},
Ronald B\"{o}hnke{\small $^{3}$},
Martin Schubert{\small $^{3}$},\\
Gilberto Berardinelli{\small $^{2}$} and 
Thomas H. Jacobsen{\small $^{4}$}\\
\vspace{5mm}
\fontsize{9}{9}\selectfont\itshape
$~^{1}$Center for Wireless Communication, University of Oulu, Finland.
$~^{2}$Department of Electronic Systems, Aalborg University, Denmark.\\
$~^{3}$Huawei Technologies Duesseldorf GmbH, Munich Research Center, Germany.
$~^{4}$Nokia Bell Labs, Aalborg, Denmark.\\
\vspace{1mm}
Emails: nurulhuda.mahmood@oulu.fi, \{rba, gb\}@es.aau.dk, \{\textit{first name.last name}\}@huawei.com, thomas.jacobsen@nokia.com\\
}

\maketitle
\thispagestyle{empty}

\begin{abstract}
The newly introduced ultra-reliable low latency communication service class in 5G New Radio depends on innovative low latency radio resource management solutions that can guarantee high reliability. Grant-free random access, where channel resources are accessed without undergoing assignment through a handshake process, is proposed in 5G New Radio as an important latency reducing solution. However, this comes at an increased likelihood of collisions resulting from uncontrolled channel access, when the same resources are preallocated to a group of users. Novel reliability enhancement techniques are therefore needed. This article provides an overview of grant-free random access in 5G New Radio focusing on the ultra-reliable low latency communication service class, and presents two reliability-enhancing solutions. The first proposes retransmissions over shared resources, whereas the second proposal incorporates grant-free transmission with non-orthogonal multiple access with overlapping transmissions being resolved through the use of advanced receivers. Both proposed solutions result in significant performance gains, in terms of reliability as well as resource efficiency. For example, the proposed non-orthogonal multiple access scheme can support a normalized load of more than $1.5$ users/slot at packet loss rates of $\sim 10^{-5} -$ a significant improvement over the maximum supported load with conventional grant-free schemes like slotted-ALOHA. 
\end{abstract}

\begin{IEEEkeywords}
URLLC, Grant-free random access, 5G NR, NOMA.
\end{IEEEkeywords}

\section{Introduction}
\label{sec:introduction}
\IEEEPARstart{U}{}ltra-reliable low latency communication (URLLC) is a new service class introduced in Fifth-generation New Radio (5G NR) cellular standard~\cite{3gppTS38300}. The reliability and latency levels offered by URLLC improve those of earlier generations of cellular standards. Examples include, isochronous real-time communication for factory and process automation in Industry 4.0 scenarios, vehicular-to-anything (V2X) communication in Automotive sector and haptic communication for tactile Internet. 

The key design challenge for URLLC is to ensure low latency and high reliability simultaneously. In the absence of a tight latency constraint, any desired level of reliability can be achieved by coding over larger blocklengths and introducing sufficient redundancy, including re-transmissions. The \textit{scheduling} and the \textit{transmission delay} are the two primary latency inducing components of a communication protocol at the lower layers that can be influenced by system design. The former is the time it takes from the point a packet arrives at the lower layer of a transmitter until it can access the channel, whereas the latter is the time it takes to successfully deliver the message. Other sources of latency include processing delays at the transmitter/receiver, propagation delay and queuing at higher layers.

The minimum scheduling unit in Long Term Evolution (LTE) is, in general, limited to the transmission time interval (TTI) of one millisecond (ms). Whereas, 5G NR has introduced the concept of `mini-slots' consisting of $1 - 13$ orthogonal frequency division multiplexed (OFDM) symbols, along with support for a scalable numerology allowing the sub-carrier spacing (SCS) to be expanded up to $240$ kHz. Collectively, this allows transmissions over shorter intervals. For example, a URLLC mini-slot of $2$ OFDM symbols at $60$~kHz SCS corresponds to a transmission time of only $35.7$~micro seconds~\cite{3gppTS38300}. 

Access to the wireless channel is generally controlled by a grant based (GB) scheduling mechanism where users attempting to access the channel have to first obtain an access grant through a four-way handshake procedure. This ensures that the user has exclusive rights to the channel, thus avoiding any potential collisions, at the expense of large latency and signalling overhead~\cite{AAS+17}. Grant-free (GF) random access, where the grant acquisition by the user prior to transmission is skipped, is proposed as a solution to reduce the access latency~\cite{berardinelli_reliabilityAnalysis_ieeeAccess2018}. 

With GF transmissions, a user with available traffic transmits the data (along with required control information) in the first transmission itself. GF transmissions can be preallocated over dedicated resources, or shared among multiple users through contention. The former is better suited for periodic traffic with a fixed pattern, whereas the latter is more resource utilization efficient and flexible, especially in case of sporadic traffic. 

GF transmissions over shared resources are subject to potential collision with other neighbouring users transmitting simultaneously, thus jeopardizing the transmission reliability. Techniques to improve the supported load with GF random access while ensuring high reliability and low latency is are therefore currently being discussed in academic research and in standardization bodies~\cite{3gppRP181477}. State of the art solutions include GF transmissions with K-repetition, where a pre-defined number of replicas are transmitted, and proactive repetition, where the transmission is proactively resent until an acknowledgment is received (also known as repetitions with early termination)~\cite{berardinelli_reliabilityAnalysis_ieeeAccess2018}. 

This article discusses GF random access in the uplink as an enabler for URLLC in 5G NR. The main contribution is two-fold, namely: \textit{i)} giving an overview of GF random access in 5G NR and discussing its shortcomings, and \textit{ii)} presenting two advanced GF schemes that go beyond 5G NR. 
In particular, the first proposal presented in Section~\ref{sub:repShared} introduces a novel transmission scheme where dedicated resources are allocated for the initial transmission, whereas blind retransmissions occur over shared radio resources. The combination of non-orthogonal multiple access (NOMA) and GF access is considered next in Section~\ref{sec:eGF_noma}. NOMA relaxes the paradigm of orthogonal transmissions by allowing different users to concurrently share the same physical resources in time, frequency, and space. More specifically, NOMA techniques are exploited at the transmitter end to improve the reliability and resource efficiency, while advanced receivers are used to resolve collisions at the receiver end.


\section{Overview of URLLC Discussions in 3GPP}

We present a brief summary of the ongoing discussion pertinent to URLLC in 3GPP Release-15, 16 and 17 in this section. The objective is to provide the reader with an overview of the important research directions in URLLC in the context of 5G NR standardization. 

The first phase of 5G NR standard covering the most basic set of use cases envisioned for 5G was completed within Release-15 and finalized in September 2018. 
The second phase of 5G NR standard, bringing the full 3GPP 5G system to its completion, is targeted by Release-16 and will be finalized at the end of 2019. Currently, there are around $25$ Release-16 study items covering a variety of topics, with a number of them involving URLLC services. Finally, Release-17 will look into emerging topics to be studied for 5G evolution systems in 2020 and onwards. 

URLLC related discussions in 3GPP NR Release-15 were primarily grouped into four different study items. The first dealt with the support of separate channel quality indicator (CQI) and modulation and coding scheme (MCS) tables for URLLC and the option of configuring two block error rate (BLER) targets for CQI reporting. The second agenda item studied the potential benefits of introducing a new Downlink Control Information (DCI) format with a smaller payload. Using a smaller DCI size permits lowering the DCI code rate, which in turn allows robust transmission on the user plane. The necessity of Physical Downlink Control Channel repetition, which can be useful in achieving high reliability in certain scenarios such as GF transmission, was investigated next. The final item was a study on handling uplink multiplexing of transmission with different reliability requirements, which considered both intra-UE and inter-UE multiplexing.

Discussions on 5G-NR URLLC in Release-16 are grouped into three different study items. The first deals with Layer-1 enhancements, including potential control channel and processing timeline improvements. The second studies the potential benefits of uplink inter-UE transmission prioritization and multiplexing. GF transmission, in particular, is enabled in Release-15 by so-called ``Configured Grant'' operations~\cite{3gppTS38300}. A study item is focused on enhancing such operations, including methods for explicit hybrid automatic repeat request acknowledgement (HARQ-ACK), to ensure K-repetitions, and mini-slot repetitions. 

URLLC studies in Release-17 will mainly focus on use cases and end-to-end performance of different applications, such as \textit{i)} audio-visual service production requiring tight synchronization and lowlatency, \textit{ii)} communication services for critical medical applications including robotic aided surgery, and \textit{iii)} support for unmanned aerial systems connectivity, identification, and tracking. The interested reader is referred to~\cite{3gppSIsite} for further details.


\section{The Basics of Grant-Free Random Access}
\label{sec:GFbasic}




The conventional GB scheduling procedure in LTE networks involves exchanging multiple messages between nodes to facilitate exclusive channel access. Due to the tight latency requirement and the associated signaling overhead, such GB schemes are not suitable for URLLC applications. GF schemes using semi-static configurations are an option to remove the signaling overhead caused by the request-followed by-grant procedure and to reduce the latency. 

The needed control information, such as time and frequency resource allocation, MCS, power control settings and HARQ related parameters, are configured by Radio Resource Control (RRC) signaling prior to the GF transmissions. In the uplink, the configured devices are connected and synchronized, thus being always ready for a URLLC transmission. The configured resources can also be shared by a number of users to increase the resource efficiency in case of sporadic traffic. However, transmissions are then susceptible to potential collisions from simultaneous transmissions of neighbouring nodes. 



This section presents a system-level simulations based performance comparison of uplink GB and several GF transmission procedures considering a large urban macro network. 

\subsection{The Considered Grant-Free Transmission Schemes}

\begin{figure}[htb]
    \centering
    \includegraphics[width=0.8\columnwidth]{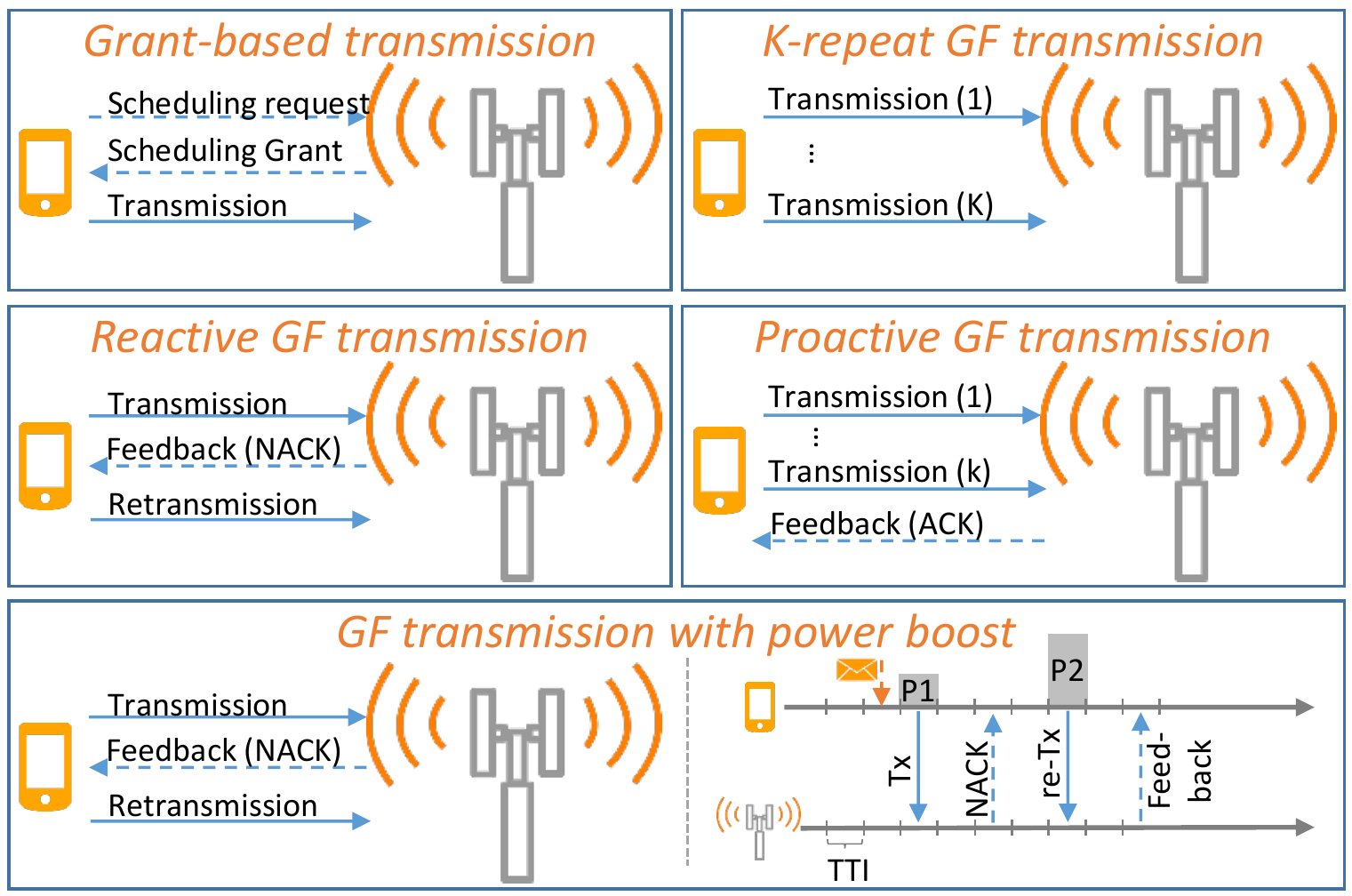}
    \caption{The analysed state-of-the-art Grant-Free Uplink HARQ Schemes for URLLC.}
    \label{fig:GFschemes}
\end{figure}

Four GF schemes are considered, namely reactive, K-repetition, proactive and GF scheme with power boost, as illustrated in Figure~\ref{fig:GFschemes}. In the reactive scheme, when the UE has finalized its initial uplink data transmissions, its signal is processed at the BS, which will then transmit a HARQ feedback (ACK/NACK). The UE re-transmits the same payload upon reception of a NACK. The time duration of the cycle from the beginning of a transmission until the processing of its feedback is called the HARQ round trip time (RTT). It is assumed that the BS spends one mini-slot for processing and one mini-slot for transmitting the feedback. 

The UE is configured to autonomously transmit the same packet \textit{K} times before waiting for a feedback from the BS in the K-repetition scheme. Each repetition can be identical, or consist of different redundancy versions of the encoded data. This method eliminates the RTT latency at the expense of potential resource wastage if the needed number of repetitions is overestimated. 

Similar to the K-repetition scheme, the UE aims at repeating the initial transmission for a number of times in the proactive scheme; with a feedback received after each transmission. This allows the UE to stop the chain of repetitions earlier in case of a positive feedback. 

The GF scheme with power boost is similar to the reactive scheme with the addition that the transmit power of each retransmission is higher than that of the previous transmission. This is motivated by the fact that HARQ retransmissions within the latency deadline should be prioritized to improve the success probability~\cite{ATB+18_pc}. Here, we discuss the applicability of improved open loop power control for GF transmissions. The considered power control is given by: $P[\text{dBm}] = \min \{ P_{max}, P_{0} + 10\log_{10}(M) + \alpha PL + g(k)\},$ where $P_{max}$ is the maximum transmit power, $P_{0}$ is the target receive power per resource block, $M$ is the number of resource blocks, $\alpha$ is the fractional path loss compensation factor, $PL$ is the path loss and the function $g(k)$ gives a power boosting step for the $k^{th}$ transmission.


\subsection{Performance Evaluation Methodology}
The simulation assumptions and parameters used for this study are in line with the guidelines for NR performance evaluations presented in~\cite{3gppTR38802}. A total of $21$ macro cells are simulated with an inter-site distance of $500$ meters, and uniformly distributed outdoor UEs. Linear minimum-mean square error with interference rejection combining (MMSE-IRC) and multi-user detection receiver is assumed in this case. A $10$ MHz band within the $4$ GHz carrier is considered. Open loop power control is used by the UE to compensate the coupling loss. The MCS is pre-configured as very conservative (QPSK with coding rate ${}^{1}/{}_{8}$), which permits the UE to transmit a $32$ byte payload in a two OFDM symbols {mini-slot} using the full band. The conventional GB scheme is considered as the baseline for comparison. The same MCS is used for the retransmissions, with Chase combining (CC) of the retransmitted packets at the receiver end. 

\subsection{Performance Results}
\begin{figure}[htb]
    \centering
    \includegraphics[width=0.8\columnwidth]{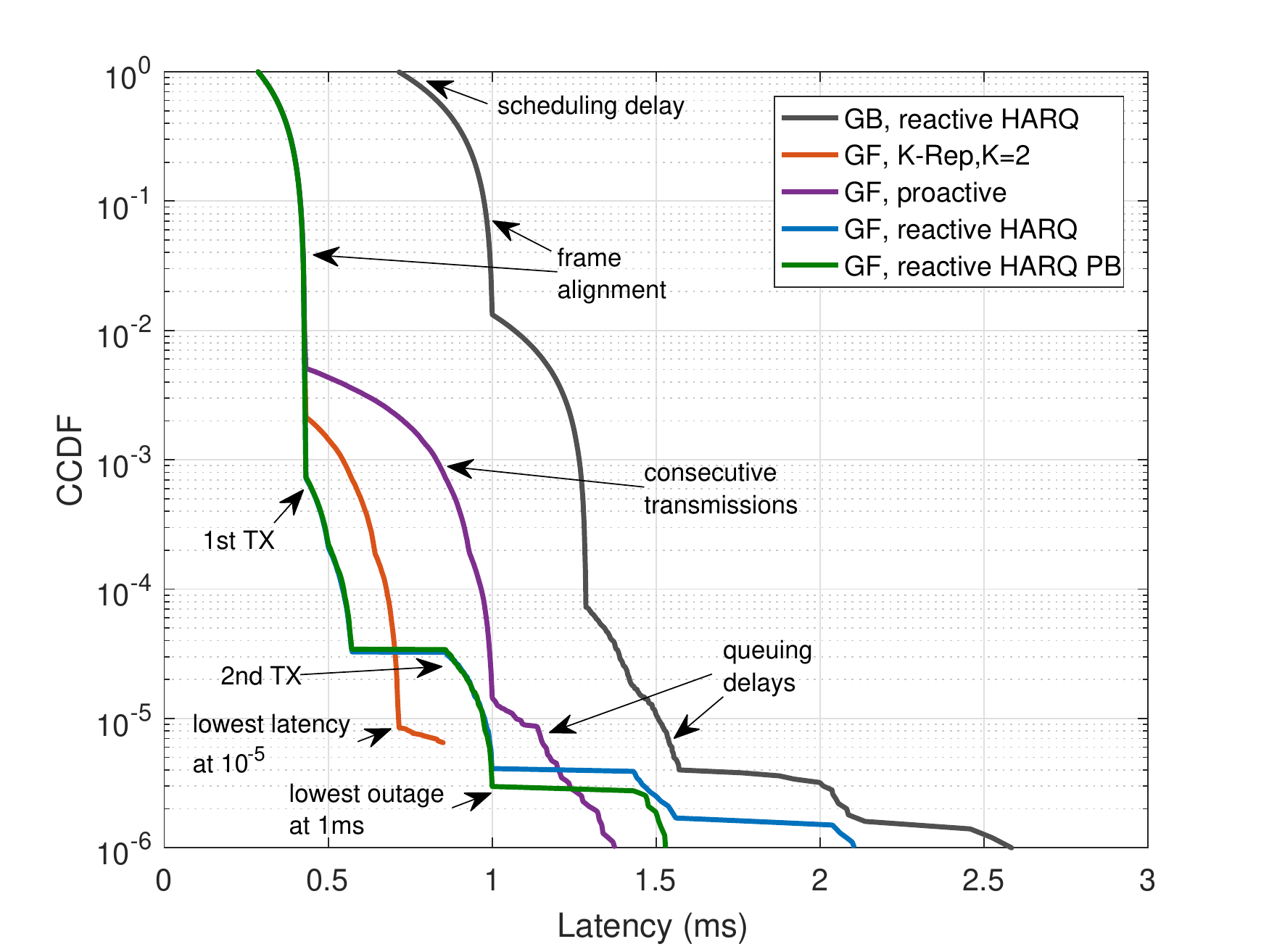}
    \caption{Performance Evaluation of Grant-Free transmission schemes: CCDF of the latency for a load of $10$ UE/cell.}
    \label{fig:perfGFcompare_ccdf}
\end{figure}

Results are presented in terms of the one-way latency for multiple payload transmissions. In Figure~\ref{fig:perfGFcompare_ccdf}, the empirical Complementary Cumulative Distribution Functions (CCDF) of the latency for the different GF transmission schemes are shown along with the baseline GB scheme at low load ($10$ UEs/cell, and Poisson arrival rate of $10$ packets per second for each UE). The horizontal axis represents the latency (in ms), whereas the vertical axis displays the outage probability. 

The GF schemes clearly provide lower latencies for the same reliability compared to the GB reference. The reactive scheme provides the best reliability for the first transmission among the GF schemes. The staircase behaviour is caused by the HARQ RTT between the retransmissions. The K-Repetition scheme with two repetitions presents similar shapes for the first and second consecutive transmissions, and is capable of providing \textit{one} ms latency at a reliability greater than $1 - 10^{-5}$ in the low load case. 

The Proactive scheme is able to achieve a very low outage performance. However, due to the feedback RTT, it is not able to terminate early before at least four consecutive transmissions, resulting in excessive resource usage. 

On the other hand, HARQ with power boost can improve the outage performance of the reactive scheme. However, the gain is limited in macro scenarios, since UEs tend to operate using maximum transmit power, thus not being able to apply the power boost in many cases.

Comparing the HARQ Reactive transmission scheme for GF and GB transmission, they show a similar staircase behaviour, with the initial step occurring at different latency and reliability combinations (e.g. $0.6$ ms and $1.6$ ms for GF and GB, respectively, where the latter is due to the scheduling procedure of GB schemes). The reason for the reliability difference for the initial transmission is due to the impact of intra-cell interference. 

We note that, despite the potential of K-Repetition to achieve the lowest latency, its achieved outage is close to the target of $10^{-5}$. This means that a higher URLLC load can not be supported without surpassing it. Conversely, the reactive HARQ scheme with possible power boosting enhancement achieves the lowest outage within the one ms latency target, and still provides room for increasing the URLLC load without violating the reliability constraint; and hence is better suited for scenarios with higher URLLC load.



\section{Enhanced GF Transmissions}
\label{sec:eGF_repetition}
\begin{figure}[htb]
    \centering
    \includegraphics[width=0.65\columnwidth]{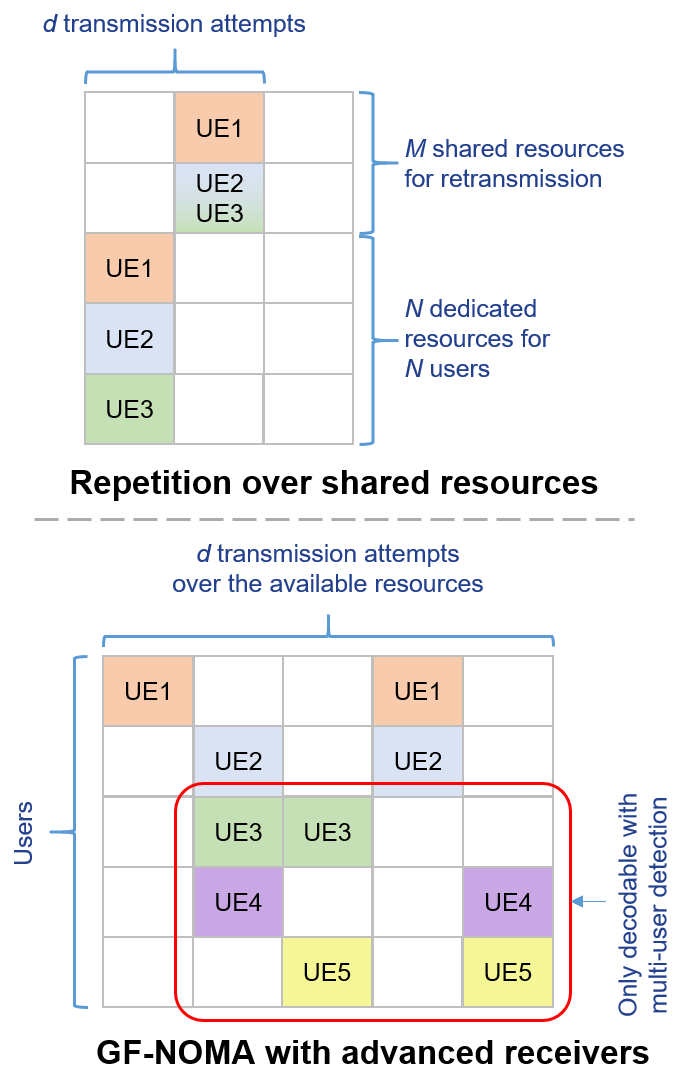}
    \caption{Schematic presentation of the proposed enhanced GF schemes with $d=2$ transmissions of each packet.}
    \label{fig:enhancedGFschemes}
\end{figure}

This section presents two novel enhanced alternatives for GF transmissions. The objective is to improve the supported URLLC load while meeting the \textit{one} ms latency target at $1- 10^{-5}$ reliability. The proposed schemes are illustrated in Figure~\ref{fig:enhancedGFschemes}. 


\subsection{Repetition with Hybrid Resource Allocation}
\label{sub:repShared}

The improved latency and reliability with the GF schemes presented in Section~\ref{sec:GFbasic} come at a cost of poor resource efficiency and poor support to higher loads due to potential collisions. Herein, we propose a hybrid scheme exploiting an advanced reception mechanism in which blind repetitions are performed with improved resource efficiency and low delay penalty.

The key idea is as follows: the initial transmission for each of the considered $N$ UEs takes place over a dedicated set of resource with a given target BLER, while subsequent retransmissions are done over a pool of shared resources of size $R$ configured by the base station. A UE can perform a total of $d$ transmission attempts using its dedicated and the shared resources, where the UE can select (randomly or according to a sequence) the retransmission resource from the available $R$ resources. 

The base station attempts to decode each transmission over the configured dedicated resources and store the successful ones. As in coded-random access (CRA) schemes~\cite{PSL+15_CRA}, we assume successive interference cancellation (SIC) at the receiver to remove the interference from decoded replicas. The initial transmissions on dedicated resources have a high success probability. Therefore, the probability that multiple non-decoded replicas collide in the shared pool is low; thus, facilitating the SIC process. 

\begin{figure}[htb]
    \centering
    \includegraphics[width=0.8\columnwidth]{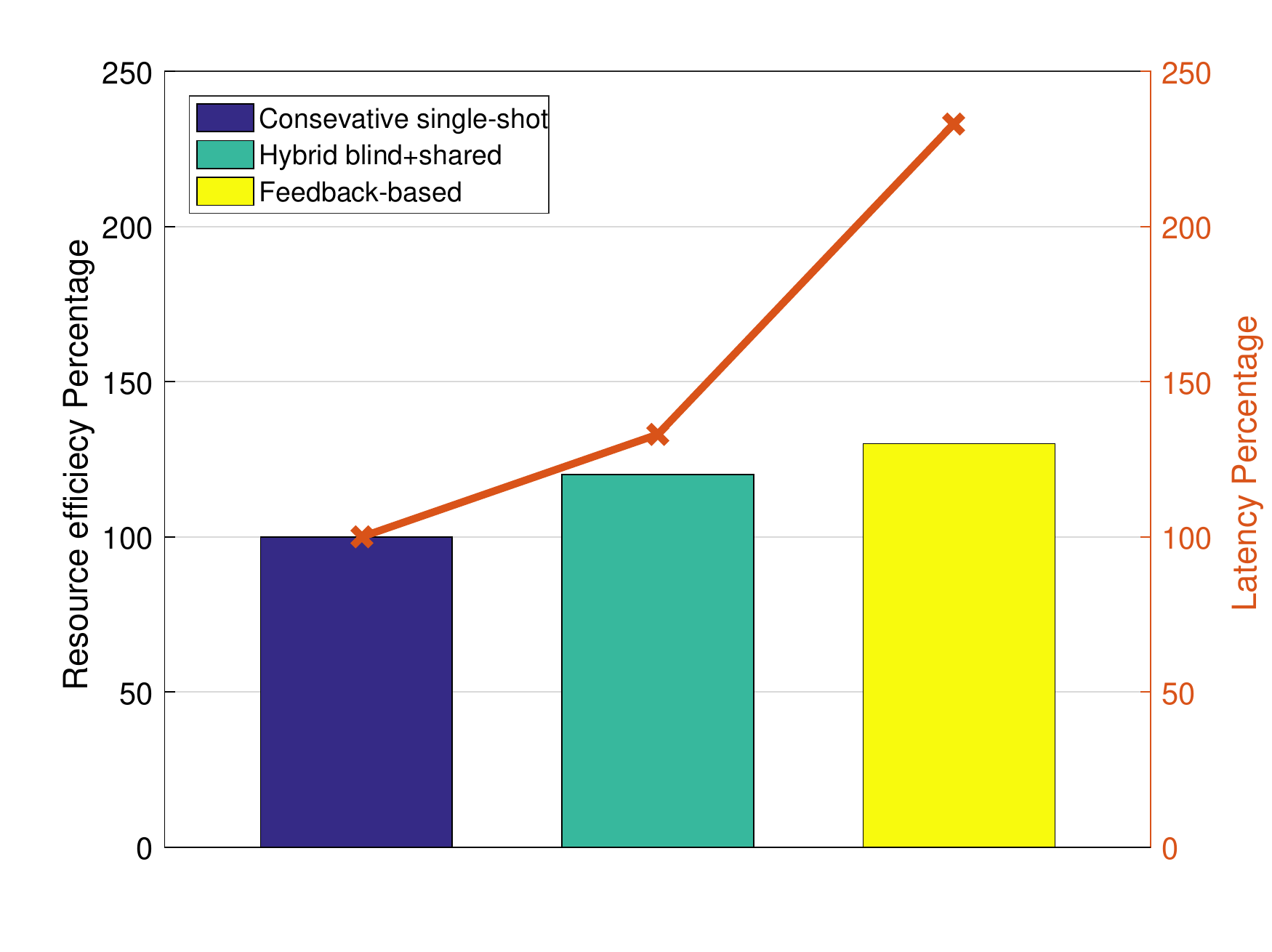}
    \caption{Comparison of resource efficient and latency performance between different schemes. Processing delay of one TTI is assumed in all cases. For the hybrid scheme, number of transmissions $d=2$, shared resource pool $R=1$, and $N=10$ users are assumed.}
    \label{fig:perfBlingRetx}
\end{figure}

The performance of the proposed scheme is evaluated for different configurations and compared against two baselines:~a robust single shot transmission, i.e. without HARQ, and a transmission including reactive HARQ with short RTT. 
A summary result is shown in Figure~\ref{fig:perfBlingRetx}. It is observed that the proposed method is nearly $23\%$ more resource efficient than single shot transmissions for a group of UEs, with a slight latency increase. Comparing with the reactive HARQ, though the resource efficiency is slightly lower, though the latency is reduced by approximately $60\%$ as it does not depend on receiving a feedback signaling. Detailed performance evaluations are furnished in~\cite{ABJ+18_blindRetx}.


\subsection{Grant-Free NOMA with Advanced Receivers}
\label{sec:eGF_noma}

The performance of classical GF access approaches like slotted ALOHA is usually limited by packet collisions. If multiple UEs transmit in the same time-frequency resource slot, the messages cannot be decoded in general. Therefore, several improved CRA schemes have been proposed recently~\cite{PSL+15_CRA}, where packets are repeated in multiple slots and collisions are resolved through SIC. The number of repetitions can be optimized for a given system load using tools from coding theory, while further gains are possible by replacing the simple packet repetitions with more general packet erasure codes. Such CRA schemes can asymptotically approach a normalized throughput of \textit{one} packet/slot for the collision channel model. 

A key assumption in most CRA studies is that packets received without collision can always be correctly decoded, though this is not always guaranteed for the transmission of short packets. On the other hand, treating packet collisions as erasures may be too pessimistic considering the possibility of employing advanced multi-user detection at the receiver. In fact, NOMA schemes, that are able to resolve interference on the physical layer based on UE-specific signatures, have attracted significant attention for 5G NR~\cite{DWY+15_noma}. 

Several NOMA candidates make use of sparse resource allocation patterns, which bears a close resemblance to CRA methods, where each active UE selects a subset of the available slots for transmission. However, NOMA allows to combine received signals from multiple slots and decode a message even if no slot is free from interference as illustrated in Figure~\ref{fig:enhancedGFschemes}. Furthermore, near-far effects resulting from differences in the path loss and/or the fading coefficients can be exploited to improve the performance of SIC.

\begin{figure}[htb]
    \centering
    \includegraphics[width=0.8\columnwidth]{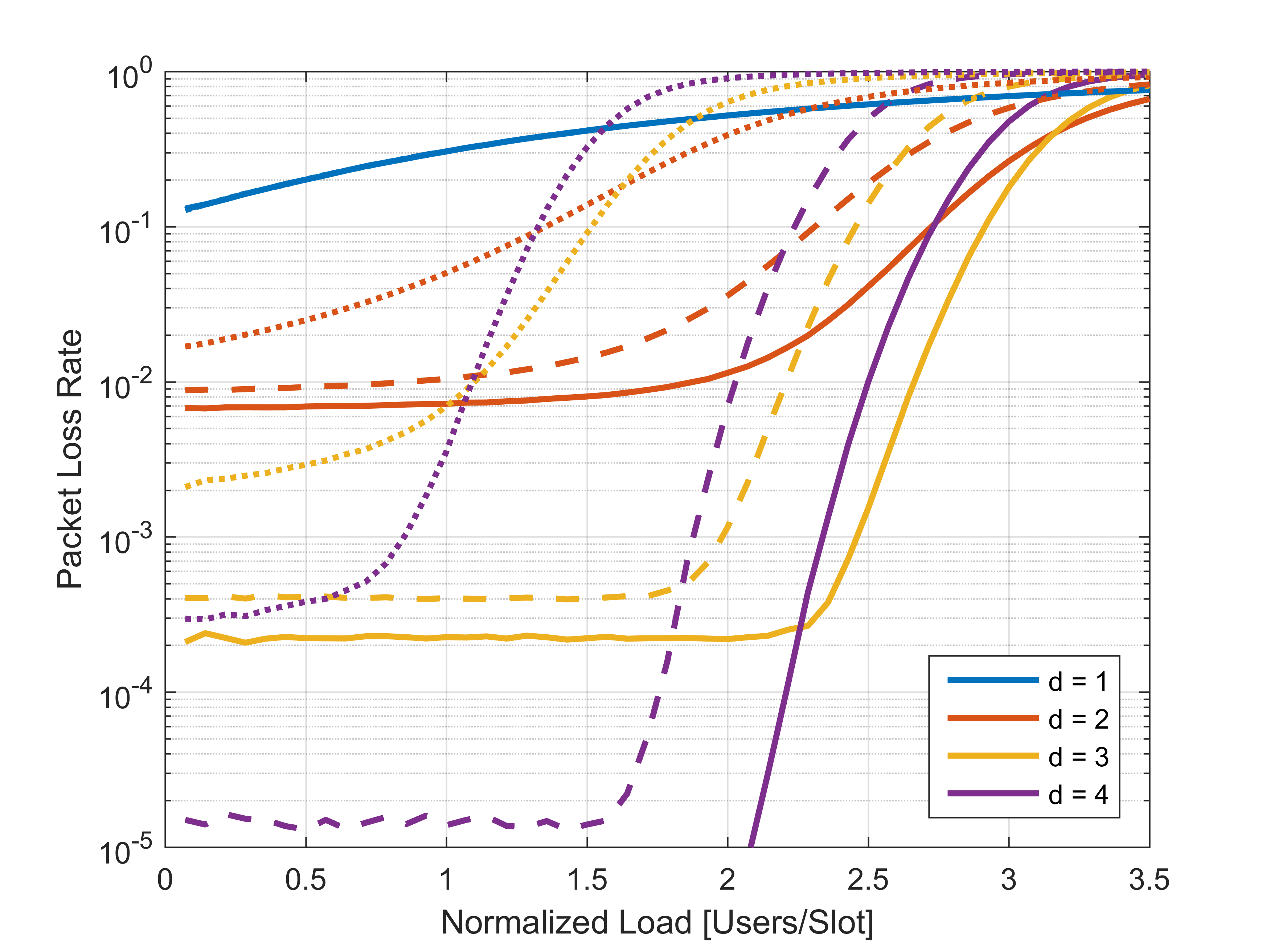}
    \caption{Comparison of grant-free access using $d$ slots for each packet and SIC with selection combining (dotted), Chase combining (dashed), and low-rate channel coding (solid).}
    \label{fig:eGF_noma}
\end{figure}

In order to illustrate the possible gains of sparse NOMA with slot combining for URLLC, we consider a scenario where each active UE selects $d$ out of $14$ slots (each comprising $240$ resource elements) to transmit a message consisting of $256$ bits. The channels are assumed to be Rayleigh block-fading with an average signal-to-noise ratio (SNR) of $9\,\text{dB}$. For simplicity, we use Polyanskiy’s normal approximation~\cite{polyanskiy_trIT2010} to determine the packet error rate and assume that interference from successfully decoded packets can be perfectly cancelled at the receiver. The following transmit and receive strategies are compared in the results presented in Figure~\ref{fig:eGF_noma}:

\begin{itemize}
\item Selection combining (dotted): Packets are repeated in $d$ slots, and the receiver selects the best slots for decoding as in conventional CRA schemes.
\item Chase combining (dashed): Packets are repeated in $d$ slots, and the receiver performs maximum ratio combining of all slots before decoding.
\item Low-rate channel coding (solid): Packets are encoded over $d$ slots, and the receiver considers all slots jointly for decoding.
\end{itemize}

The special case $d=1$ corresponds to slotted ALOHA. Transmitting a message in multiple slots provides a diversity gain regarding both the strength of the desired signal and the interference. However, selecting only the best slots for decoding as in CRA does not make use of all available information at the receiver and fails to meet the strict reliability requirements of URLLC, even for small system loads. 

On the other hand, Chase combining repeated packets yields a significant performance improvement. For the considered example, the target packet loss rate of $10^{-5}$ is just missed with $d = 4$ repetitions. The additional coding gain of the third method based on low-rate channel coding enables URLLC with GF access even for highly overloaded systems with \textit{two} packets/slot, which means that on average eight users are transmitting simultaneously in each slot for $d = 4$. Note that the curves exhibit a threshold behavior, which was also observed for the fundamental limits in \cite{KP19_MAClimits}: the packet loss rate remains almost constant up to a certain load and then sharply increases, which means that the interference can be completely removed with high probability up to this point. Increasing $d$ further reduces the error floor, but also the load threshold.

The results clearly indicate the benefits of sparse NOMA compared to simple CRA schemes with slot-wise decoding, albeit at the expense of increased processing complexity at the receiver. Further gains are possible, e.g., by using more advanced joint multi-user detection algorithms and multiple receive antennas to separate the superimposed signals. Note that user activity detection and channel estimation are required before the data detection, but this may also be done jointly \cite{DDZ+18_jcemud}. For block-fading channels, there is further a tradeoff between the diversity achieved by using multiple slots and the required channel estimation overhead \cite{ODS+18_PilotVsNoncoherent}. These practical issues are facilitated by the sparse resource allocation.


\subsection{Key Take-away Message}
\label{sub:eGF_summary}

The two schemes are motivated by the fact that different URLLC use cases can have very different traffic profiles. The first scheme discussed in Section~\ref{sub:repShared} is particularly well suited for periodic URLLC messages, as in Industry 4.0 scenarios~\cite{BML+18_wirt}. Orthogonal resources can be allocated in a semi-persistent manner for the first transmission, followed by retransmission over shared resources to ensure very high reliability. On the other hand, the GF-NOMA scheme with advanced receivers proposed in Section~\ref{sec:eGF_noma} rather targets random packet arrivals, e.g. emergency messages in V2X scenarios. 

The key take-away message in both cases is that packet collisions over shared resources are not bad \textit{per se}. Resolving them with advanced receiver design instead of treating the resulting interference as noise leads to higher reliability, lower latency and improved resource efficiency.


\section{Overview of the Proposed Grant-Free Access Solutions and Outlook}
\label{sec:overview}

\subsection{Overview}
An overview of the discussed GF access solutions is shown in Table~\ref{tab:solution_overview}, along with their link to ongoing/upcoming URLLC standardization effort in 3GPP. 


\begin{table*}[!t]
\centering
\caption{Overview of existing and proposed GF access solutions}
\label{tab:solution_overview}
\renewcommand{\arraystretch}{1.3}
\begin{tabular}{l p{2.5cm} p{5cm} p{5cm} p{2.5cm}}
\toprule
 & \textbf{Scheme} & \textbf{Description} & \textbf{Performance} & \textbf{Standardization Status}\\
\midrule
\multirow{2}[25]{*}{\rotatebox{90}{\textbf{Standard GF schemes}}} 
& \textit{Reactive GF }& 
\vspace{-10pt}
\begin{itemize}
\item GF initial transmission
\item Retransmission upon NACK
\end{itemize}
& 
\vspace{-10pt}
\begin{itemize}
\item Good performance at moderate loads
\item Long latency tail 
\end{itemize}
& Supported in Release-15\\ \cmidrule{2-5}
& \textit{K-Repetitions GF} & 
GF transmission repeated $K$ times
& 
\vspace{-10pt}
\begin{itemize}
\item Lowest latency at low loads
\item Performance degrades quickly with increase in load 
\item Resource in-efficient due to blind retransmissions
\end{itemize}
& Supported in Release-15\\ \midrule
\multirow{2}[30]{*}{\rotatebox{90}{\textbf{Non-standard GF schemes}}}
& \textit{Proactive GF} & 
GF transmissions repeated until ACK received & 
\vspace{-10pt}
\begin{itemize}
\item Similar performance trend as K-Repetition at low loads
\item Performance degradation with increase in load is lower
\item More resource efficient than K-Repetition with high $K>4$
\end{itemize}
& Not specified in Release-15, but implementable\\ \cmidrule{2-5}
& \textit{Reactive GF with power boost} & 
Reactive GF with transmit power boost in retransmissions & 
\vspace{-10pt}
\begin{itemize}
\item Power boost leads to higher reliability 
\item More resource efficient as multiple retransmissions are avoided
\end{itemize}
& Not standardized but can be implemented with Release-15 \\ \midrule
\multirow{2}[30]{*}{\rotatebox{90}{\textbf{Enhanced GF schemes}}} 
& \textit{Repetition with hybrid allocations}& 
\vspace{-10pt}
\begin{itemize}
\item GF initial transmission over dedicated resource
\item $d$ retransmission over a shared resource pool
\end{itemize}
& 
\vspace{-10pt}
\begin{itemize}
\item $\sim 60\%$ lower latency compared to reactive GF scheme
\item $23\%$ efficiency gain over robust single-shot transmission
\end{itemize}
& Under investigation in Release-16\\ \cmidrule{2-5}
& \textit{GF NOMA with advanced receivers} & 
\vspace{-10pt}
\begin{itemize}
\item GF transmissions over multiple slots
\item Considers low-rate coding across the multiple transmission block
\item CC at the receiver end
\end{itemize}
& 
\vspace{-10pt}
\begin{itemize}
\item CC can support close to $1$ UE/slot at very low outage
\item The performance is further improved with joint decoding
\end{itemize}
& NOMA schemes are being studied for Release-16\\ 
\bottomrule
\end{tabular} 

\end{table*}
%

\subsection{Conclusions and Future Outlook}
Enabling resource-efficient wireless communication with low latency and high reliability is a key design goal for URLLC. Grant-free transmissions, where users initiate a transmission without a scheduling grant from the network, has emerged as a promising latency reducing solution. However, the lack of admission control makes it vulnerable to potential collisions from other co-existing users, jeopardizing the reliability. This necessitates enhanced GF schemes that can ensure high reliability without sacrificing efficiency.

This article presents GF random access schemes as a promising URLLC solution for 5G NR. We have presented and evaluated GF transmissions in details, followed by proposals of two enhanced resource efficient GF schemes that improve the reliability and latency performance while maintaining a high resource efficiency. The scheme with repetition over shared resources leads to halving the transmission latency as compared to legacy feedback based schemes, whereas the proposed GF transmission with advanced receiver allows to operate even in overloaded systems at outage probabilities as low as $\sim 10^{-5} -$ thereby significantly boosting the supported URLLC load.

\section*{Acknowledgements}
This work has partly been performed in the framework of the Horizon 2020 project ONE-5G (ICT-760809) receiving funds from the European Union, and partly under the Academy of Finland 6Genesis Flagship (grant no. 318927). The authors would like to acknowledge the contributions of their colleagues in the project, although the views expressed in this work are those of the authors and do not necessarily represent the project.



\begin{IEEEbiographynophoto}{Nurul Huda Mahmood} was born in Chittagong, Bangladesh. He is currently a senior research fellow at Centre for Wireless Communications, University of Oulu, Finland, where he is involved in the \href{https://www.oulu.fi/6gflagship/}{\em 6G Flagship} program. His research interests include resource optimization techniques with focus on URLLC/MTC, and modeling and performance analysis of wireless communication systems.
\end{IEEEbiographynophoto}

\begin{IEEEbiographynophoto}{Renato Abreu} received the M.Sc. degree in electrical engineering from the Federal University of Amazonas, Brazil, in 2014. From 2008 to 2016, he was with the former Nokia Technology Institute (INDT), Brazil, developing industrial test systems for mobile devices and researching wireless systems. Since 2016, he has been a Ph.D. fellow with Wireless Communication Networks Section (WCN), Aalborg University (AAU), Denmark, where he works in cooperation with Nokia Bell Labs. His research focus is on solutions for URLLC in 5G NR.
\end{IEEEbiographynophoto}

\begin{IEEEbiographynophoto}{Ronald B\"{o}hnke} received his Dipl.-Ing. and Dr.-Ing. degrees in electrical engineering from the University of Bremen, Germany, in 2002 and 2014, respectively. From 2002 to 2010, he worked as a research and teaching assistant in the Department of Communications Engineering at the University of Bremen. In 2002, he was a visiting researcher at the Fraunhofer Heinrich Hertz Institute, Berlin. From 2010 to 2014, he was with the Institute for Communications and Navigation and the Institute for Communications Engineering at the Technische Universität München (TUM), Germany. In 2014, he joined the Huawei German Research Center in Munich, Germany. He has been involved in several national and EU funded research projects. His main research interests include efficient detection algorithms and adaptive transmission for MIMO systems, non-orthogonal multiple access, channel coding and modulation.
\end{IEEEbiographynophoto}

\begin{IEEEbiographynophoto}{Martin Schubert} received his doctoral degree in electrical engineering from the Technische Universit\"{a}t, Berlin, Germany, in 2002. From 2003-2012, he has been with the Fraunhofer Heinrich Hertz Institute (HHI) where he worked as senior researcher, lecturer, and team leader. He was a corecipient of the VDE Johann-Philipp-Reis Award in 2007, and he co-authored the 2007 Best Paper Award of the IEEE Signal Processing Society. From 2009-2013 he has been Associate Editor of the IEEE Transactions of Signal Processing. Dr. Schubert is Senior Member of the IEEE. Since 2013 he is Principal Researcher at the Huawei Munich Research Center, where he is working on radio interface designs.
\end{IEEEbiographynophoto}

\begin{IEEEbiographynophoto}{Gilberto Berardinelli} received the bachelor's and master's degrees (cum laude) in telecommunication engineering from the University of L'Aquila, Italy, in 2003 and 2005, respectively, and the Ph.D. degree from AAU, Denmark, in 2010. He is currently an Associate Professor with WCN, AAU. He also works in close cooperation with Nokia Bell Labs. He is currently part of the EU funded research project ONE5G that focuses on end-to-end-aware optimizations and advancements for the network edge of 5G NR. His research interests are mostly focused on physical layer, medium access control, and radio resource management design for 5G systems.
\end{IEEEbiographynophoto}

\begin{IEEEbiographynophoto}{Thomas H. Jacobsen} received his M.Sc. in engineering (network and distributed systems) from Aalborg University, Denmark, in 2015. He is currently with Nokia Bell Labs as a device standardization research expert, while pursuing a Ph.D. at Aalborg University, Denmark. His research focus on Industrial IoT related topics and radio resource mechanisms for uplink grant-free URLLC.
\end{IEEEbiographynophoto}

\end{document}